\documentstyle[aps]{revtex}
\input{psfig}

\begin{document}

\textheight 8.8in
\textwidth 6.5in
\topmargin -.25in
\oddsidemargin -.25in
\evensidemargin 0in
\baselineskip 14pt
\def\hm{\ \rm {\it h}^{-1} Mpc}

\title{Extended Quintessence}

\author{Francesca Perrotta$^{1}$\footnote{perrotta@sissa.it}, 
Carlo Baccigalupi$^{1}$\footnote{bacci@sissa.it}, Sabino
Matarrese $^{2}$\footnote{matarrese@pd.infn.it}}
\address{$^{1}$ SISSA/ISAS, Via Beirut 4, 34014 Trieste, Italy;\\
$^{2}$ Dipartimento di Fisica `Galileo Galilei', Universit\'a di Padova,\\ 
and INFN, Sezione di Padova, Via Marzolo 8, 35131 Padova, Italy.}

\baselineskip 10pt
\maketitle
\begin{abstract} 
We study Quintessence cosmologies in the context of scalar-tensor theories 
of gravity, where a scalar field $\phi$, assumed to provide 
most of the cosmic energy density today, 
is non-minimally coupled to the Ricci curvature 
scalar $R$. Such `Extended Quintessence' cosmologies have 
the appealing feature that the same field causing the time 
(and space) variation of the cosmological constant is   
the source of a varying Newton's constant 
{\em \`a la} Jordan-Brans-Dicke. 
We investigate here two classes of models,  
where the gravitational sector of the Lagrangian is $F(\phi )R$ 
with $F(\phi )=\xi\phi^{2}$ (Induced Gravity, IG) and 
$F(\phi )=1+\xi\phi^{2}$ (Non-Minimal Coupling, NMC).  
As a first application of this idea we consider a specific model, 
where the Quintessence field, $\phi$, obeying the simplest inverse power 
potential, has $\Omega_{\phi}=0.6$ today, 
in the context of the Cold Dark Matter scenario for structure formation
in the Universe,
with scale-invariant adiabatic initial perturbations.
We find that, if $\xi\lesssim 5\times 10^{-4}$ for IG 
and $\xi\lesssim 5\times 10^{-3}(\sqrt{G}\phi_{0})^{-1}$ for NMC 
($\phi_{0}$ is the present Quintessence value) 
our Quintessence field satisfies the existing solar 
system experimental constraints. 
Using linear perturbation theory we then obtain 
the polarization and temperature anisotropy spectra of the 
Cosmic Microwave Background (CMB) as well as the matter 
power-spectrum. The perturbation behavior possesses distinctive 
features, that we name `QR-effects': 
the effective potential arising from the coupling with $R$ 
adds to the true scalar field potential, altering the cosmic 
equation of state 
and enhancing the Integrated Sachs-Wolfe effect. 
As a consequence, part of the CMB anisotropy level on COBE scales
is due to the latter effect, and the cosmological perturbation 
amplitude on smaller scales, including the oscillating region of 
the CMB spectrum, has reduced power; this effect is 
evident on CMB polarization and temperature fluctuations, as well as 
on the matter power-spectrum today. Moreover, the acoustic peaks and the 
spectrum turnover are displaced to smaller scales, 
compared to ordinary Quintessence models, because of 
the faster growth of the Hubble length, 
which - for a fixed value today - 
delays the horizon crossing of scales larger than the horizon
wavelength at matter-radiation equality 
and slightly decreases the amplitude of the acoustic oscillations. 
These features could be detected in the upcoming observations 
on CMB and large-scale structure. 
\end{abstract}

\section{Introduction}

Recently a lot of work focused on the cosmological role of a
minimally-coupled scalar field, considered as a "Quintessence" (Q) component
which is supposed to provide the dominant contribution to the energy density 
of the Universe today in the form of dynamical vacuum energy or `decaying
cosmological constant' \cite{Stain1} -\cite{CF}. 
This work was motivated by the  observational trend for an
accelerating Universe, as suggested by distance measurements to type Ia
Supernovae (see e.g. \cite{Perlm}, \cite{garna}). \\
The main feature of such a vacuum energy component, that could
also allow to  
distinguish it from a cosmological constant, is its time-dependence and 
the wider range of possibilities for its equation of state compared to 
the cosmological constant case.
In order not to violate the principle of general covariance, 
such a time varying scalar field should also develop spatial
perturbations.   
Noticeably, if one assumes that the Universe has critical density
as predicted by most inflationary models, this component could be the form 
in which nearly two thirds of such a density resides. \\ 
The major success of the Quintessence models is their capability to 
offer a valid alternative explanation of 
the smallness of the present vacuum energy density 
instead of the cosmological constant; indeed, we must have 
$| \rho_{vac}| < 10^{-47}$ GeV$^4$ \ today, while
quantum field theories would predict a value for the cosmological constant 
energy density which is larger by more than 100 orders of magnitude (for a
review, see for example \cite{Lambda},\cite{Sahni}). 
On the other hand, in all the models considered up to now, the vacuum  
energy associated to the Quintessence is dynamically evolving towards
zero driven by the evolution of the scalar field. \\
Furthermore, in the Quintessence scenarios one can select a subclass of
models, which admit "tracking solutions" \cite{track}: here a
given amount 
of scalar field energy density today can be reached starting from a  wide 
set of initial conditions. 
We are therefore encouraged to pursue the investigation of Quintessence
models.\\    
The classical tests of gravity theories put severe constraints on the
scalar field term arising in the action; by far, the strongest constraint 
being the E\"{o}tv\"{o}s-Dicke experiment \cite{Roll}. 
To avoid having to require a 
coincidental similarity between different Yukawa couplings, one must
constrain to very small values any explicit coupling of the scalar field
to ordinary matter \cite{Ratra}. 

The possible coupling between a Quintessence field and light matter 
has been explored in \cite{carroll} and it is subject to restrictions 
from the constraints on the time variation of the constants of 
nature; a recent work explores several cosmological consequences 
of a coupling between Quintessence and matter fields \cite{ame0}. 
Moreover, a possible coupling between the scalar field, 
modeling the Quintessence component, and the Ricci scalar $R$ 
is not to be excluded in the context of generalized 
Einstein gravity theories. 
Due to the required flatness of its 
potential to achieve slow-rolling, the coupling between 
Quintessence field and other physical entities 
gives rise to long-range ($>H_0^{-1}$) interactions; 
in the case of coupling with the Ricci scalar, these long-range 
interactions are of gravitational nature, giving rise to time variation 
of the Newtonian constant, so that the 
coupling parameter is constrained by solar system experiments 
\cite{Chiba}. Recently,  
some authors \cite{Ame}, \cite{Uzan} considered scalar-tensor theories of 
gravity in the context of Quintessence models, 
studying the existence and stability of cosmological scaling solutions.\\

Here we present the evolution of cosmological perturbations in some
subclass of these theories, where the scalar field coupled with $R$ 
will be proposed as the Quintessence candidate, and we discuss its role  
on CMB anisotropies and on structure formation in the Universe.\\ 
We name our model `Extended Quintessence' (EQ), in analogy with 
Extended Inflation models \cite{exte}, where a Jordan-Brans-Dicke (JBD)
scalar field \cite{JBD} was added to the action to solve the `graceful
exit' problem of 
`Old Inflation'. Of course, the similarity 
is not complete: in Extended Inflation a second scalar field -
the `Inflaton', undergoing a first-order phase transition, was the actual 
source of vacuum energy during inflation. Here, instead, we are supposing 
that our non-minimally coupled scalar field has its own potential which 
gives rise to a time (and space) varying cosmological constant term  
dominating the present-day energy density of the Universe. \\
The first proposal of using a non-minimally coupled scalar field to 
obtain a decaying cosmological constant dates back to 1983, when 
Dolgov \cite{dolgov} suggested to exploit the effective negative energy term 
contributed by the coupling of a massless scalar field 
with the Ricci scalar $R$ 
to drive the overall vacuum energy density to zero
asymptotically. The main problem with such a simple model is that the
interesting 
dynamical range is achieved when the change in the effective Newton's 
constant strongly contradicts upper limits on solar system experiments
(see \cite{Sahni}). 
Our model will differ from Dolgov's idea in that  
we will not assume that the non-minimal coupling term is the only cause of 
time variation for the effective vacuum energy contribution. 
This allows us to easily achieve consistency  
with the solar system experimental limits on the coupling constants. 

In this paper we present the background and perturbations equations 
in the most general form and we consider their evolution for 
Induced Gravity (IG) and Non-Minimally Coupled (NMC) scalar field models. \\
The Induced Gravity model was initially proposed by Zee in 1979
\cite{Zee}, as a theory for the gravitational interaction incorporating 
the concept of spontaneous symmetry breaking; it was based on the observation 
in gauge theories that dimensional coupling constants arising in a 
low-energy effective theory can be expressed in terms of vacuum expectation 
values of scalar fields.
This model was subsequently incorporated in  models of 
inflation with a slow-rolling scalar field \cite{Accetta}; 
in a modified form it was 
the key ingredient of the Extended Inflation \cite{exte} class of models.  
More recently, it has also been adopted in open inflation 
models \cite{Bellido}. 
In \cite{Zee}, a scalar field coupled to gravity by a term proportional
to $R \phi^2$ in the Lagrangian, is anchored by a symmetry-breaking
potential to a fixed value which eliminates the potential energy in
the present broken-symmetric phase of the world. We propose here a
different role for this scalar field, in the sense that we keep the same
coupling with the Ricci scalar as \cite{Zee}, but we allow for a larger
class of potentials than the Coleman-Weinberg one, also including
potentials that do not possess a minimum and can therefore contribute to
the present Quintessence energy density. \\
The second class of theories to which we apply our treatment is that of
non-minimal coupling of a scalar field to the Ricci curvature, described
extensively in curved space quantum field theory textbooks (e.g.
\cite{Birrel}). 

The work is organized as follows: in Sec. II we present 
the relevant equations, defining the dynamical system 
for the background as well as for the perturbations in non-minimally 
coupled scalar field cosmologies; Sec. III is devoted to the definition 
of the IG and NMC models and to the analysis of the background 
evolution; Sec. IV contains and discusses the results of 
the numerical integration. Finally, Sec. V contains 
a brief summary of the results and some concluding remarks. 

\section{Cosmological equations in scalar-tensor theories of gravity}

Our purpose is to describe a class of scalar-tensor theories of gravity
represented by the action 
\begin{equation}
\label{action} 
S=\int  d^4 x {\sqrt{-g} \over \beta} \left[ {1 \over 2} f(\phi, R) -
 {1 \over 2} \omega ( \phi) \phi^{; \mu} \phi_{; \mu} -V( \phi)
+\beta L_{fluid} \right] \;,
\end{equation} 
where $R$ is the Ricci scalar, $\beta$ is a constant needed to fix units  
and $L_{fluid}$ is a classical
multicomponent-fluid Lagrangian including also minimally coupled scalar
fields, if any. We disregard any possible coupling of our scalar field
with ordinary matter, radiation and dark matter \cite{dmq}. \\

We assume a standard Friedman-Robertson-Walker (FRW) form for the
unperturbed background metric and we restrict ourselves to a spatially 
flat universe. \\
We are using units where $c \equiv 1$, but the convention concerning $8
\pi G$ will be stated later, since it will depend on the choice of a
specific theory included in this general description. Instead, following
\cite{HW2}, we will choose the relation $G_{ \mu \nu} = T_{ \mu \nu}$ to 
identify $T_{ \mu \nu}$. Greek indices will be used for space-time 
coordinates, latin ones will label spatial ones. We use the 
signature $(-,+,+,+)$.   
By defining $F \equiv \partial
f / \partial R$, the gravitational field equations derived by the action
(\ref{action}) are 
\begin{equation}
\label{G}
G_{\mu \nu} = T_{\mu \nu} \equiv {1 \over F } \left[ \beta 
T^{fluid}_{\mu \nu} + \omega \left( \phi_{, \mu} \phi_{, \nu}
-{1 \over 2 } g_{\mu \nu} \phi_{, \sigma} \phi^{; \sigma} \right ) +
g_{\mu \nu} { f- RF-2V \over  2} + F_{, \mu ; \nu} - g_{\mu \nu} 
F^{;\sigma}_{;\sigma} \right]
\end{equation} 
Here $G_{\mu \nu}$ is the Einstein tensor, and all the other contributions
have been absorbed in $T_{\mu\nu}$; 
as noted in (\cite{HW1},\cite{HW2}), if one writes the gravitational
field equation in this form, then $T_{\mu \nu}$ can be treated as an
effective stress-energy tensor, which allows to use the standard Einstein 
equations by simply replacing the fluid quantities with the effective
ones. The background effective quantities following from the
definition of $T_{\mu \nu}$ are 
\begin{equation}
\label{rhop}
 \rho={ 1 \over F} \left(  \beta \rho_{fluid} + {\omega \over 2 a^2}
\dot{\phi}^2 
 +{RF - f \over 2} +V- {3 {\cal H} \dot{F} \over a^2}   \right)
 \  ; \ \ \
  p={ 1 \over F } \left(\beta p_{fluid}+  {\omega \over 2 a^2}
\dot{\phi}^2
- {RF - f \over 2}-V+ {\ddot{F} \over a^2} +
{{\cal H}\dot{F} \over  a^2}  \right) \; ,
\end{equation}
where the overdot denotes differentiation with respect to the conformal time
$\tau$ and ${\cal H } = {\dot a}/a$. 

The background FRW equations read 
\begin{equation} 
\label{Friedmann}
{\cal H}^2= {1 \over 3F} \left(
a^2 \beta {\rho}_{fluid} +{\omega \over 2} \dot{\phi}^2 + {a^2 \over 2}
(RF-f)
+a^2 V - 3 {\cal H} \dot{F} \right) \;,
\end{equation}
\begin{equation}
\label{Friedmann2} 
\dot{\cal H}={\cal H}^2- {1 \over 2F} \left(
a^2\beta ({\rho}_{fluid}+p_{fluid})
+\omega \dot{\phi}^2 + \ddot{F} -2 {\cal H} \dot{F}  \right)
\end{equation}
while the Klein-Gordon equation reads 
\begin{equation}
\label{KG}
\ddot{\phi}+2{\cal H} \dot{\phi}=-{1 \over 2 \omega} \left(
 \omega_{, \phi} \dot{\phi}^2 - a^2 f_{, \phi} +2 a^2 V_{, \phi} \right) \;.
\end{equation}
Furthermore, the continuity equations for the 
individual fluid components are not directly affected by the
changes in the gravitational field equation, and for the $i-$th component
\begin{equation}
\dot{\rho}_i = -3 {\cal H} ({\rho}_i + p_i)\ .
\end{equation}

In this background, the trace of (\ref{G}) becomes 
\begin{equation}
\label{Ricci}
-R={1 \over F} \left[ 
\beta T_{fluid} + 
\omega { \dot{\phi}^2 \over a^2 } + 2(f-RF-2V)+3\left( {\ddot{F} \over
a^2}+ 2 { {\cal H} \dot{F} \over a^2} \right)
 \right]\ ,
\end{equation}
recalling that $T_{fluid}=-\rho_{fluid}+3p_{fluid}$; 
note that $R$ also appears in the right hand side of the equation, 
unless $f$ is of the form $f(\phi ,R)=F(\phi)R$. 
An expression that will be useful in the following is that of the 
Ricci scalar, 
\begin{equation}
R={6 \over a^2} ( {\dot{\cal H}}^2+  {\cal H}^2)\ .
\label{riccih}
\end{equation}
Our treatment of the perturbations to this background follows (and
generalizes) a previous work \cite{CF}, based on the formalism 
developed in \cite{MB} to 
describe the evolution of perturbations in the synchronous gauge. \\
A scalar-type metric perturbation in the synchronous gauge is
parameterized as 
\begin{equation}
ds^2=a^2 [-d\tau ^2 + (\delta_{i j }+h_{i j })dx^i dx^j ] \  \ ,  
\end{equation}
\begin{equation}
h_{i j}( {\bf x}, \tau ) = \int d^3 k e^{i {\bf k} \cdot {\bf x}}
\left[ 
{\bf {\hat{k}_i  \hat{k}_j}}  h( {\bf k}, \tau)
+ ( {\bf {\hat{k}_i  \hat{k}_j}}- {1 \over 3} \delta_{ i j} )
6 \eta  ( {\bf k}, \tau)\right]\ ,
\end{equation}
where $h$ denotes the trace of $h_{ij}$; the fluid perturbations 
are described in terms of the variables 
$\delta \rho = - \delta T_{0}^{0}$, $\delta p = 
\delta T_{i}^{i}/3$, $(p+\rho)\theta= ik^{j}\delta T_{j}^{0}$
and $(p+\rho)\sigma=-\left(\hat{\bf k}_{i}\hat{\bf k}_{j}-{1\over 3}
\delta_{ij}\right)\Sigma_{j}^{i}$. 

In terms of the effective fluid, the perturbed quantities can be written as
\begin{equation}
\label{deltarho}
\delta \rho = {1 \over F } \left[
\beta \delta {\rho}_{fluid} + \omega {\dot{\phi} \delta\dot{\phi} \over a^2} 
+{1 \over 2} ({ \dot{\phi}^2 {\omega}_{,\phi} \over  a^2}
-f_{,\phi}+2V_{,\phi}) \delta \phi
-3{ {\cal H} \delta \dot{F} \over a^2} -
\left(  { \rho + 3 p \over 2 } + {k^2 \over a^2} \right) \delta F +
{\dot{F}\dot{h} \over 6a^{2}}      \right] 
\end{equation}

\begin{equation}
\label{deltap}
\delta p = {1 \over F } \left[ \beta \delta p_{fluid} +
 \omega {\dot{\phi} \delta\dot{\phi} \over a^2}
+ {1 \over 2} ({ \dot{\phi}^2 {\omega}_{,\phi} \over  a^2}
+f_{,\phi}-2V_{,\phi}) \delta \phi
+{\delta \ddot{F}\over a^{2}} + { {\cal H} \delta \dot{F} \over a^2} +
\left(  {p- \rho  \over 2 } + {2k^2 \over 3a^2} \right) \delta F 
-{1 \over 9 } {\dot {F} \dot{h}  \over a^2}       \right]
\end{equation}

\begin{equation}
\label{theta}
(p+ \rho) \theta = {\beta (p_{fluid}+ \rho_{fluid} ) \theta_{fluid}  \over
F }
- {k^2 \over a^2 } \left( 
{ - \omega \dot{\phi} \delta \phi - \delta\dot{F } +{\cal H} \delta
F \over F} \right)
\end{equation}

\begin{equation}
\label{shear}
(p+ \rho) \sigma=  {\beta (p_{fluid}+ \rho_{fluid} ) \sigma_{fluid}  \over
F }
+ {2k^2 \over 3a^2 F}  
 \left(    \delta F + 3 { \dot{F} \over k^2} ( \dot{\eta} + {\dot{h} 
\over 6} )  \right)  \;. 
\end{equation}

The perturbed Klein-Gordon equation reads
\begin{equation}
\label{KGpert}
\delta \ddot{\phi} + \left( 3 {\cal H} + {\omega_{,\phi} \over \omega}
\dot{\phi} \right)\delta\dot{\phi} + \left[ k^2 + {\left( {\omega_{,\phi}
\over \omega} \right) }_{,\phi} {\dot{\phi}^{2} \over 2 } +
a^2 {\left( {-f_{, \phi} +2V_{, \phi} \over 2 \omega } \right) }_{,\phi}
\right] \delta \phi=  {\dot{\phi} \dot{h} \over 6 } + {a^2 \over 2 \omega}
f_{, \phi  R} \delta R \ . 
\end{equation}
Note the presence of the Ricci curvature scalar $R$ in the 
$f_{, \phi}$ term in the left hand side, as well as its perturbation 
$\delta R$ in the right hand one. 

All these ingredients have to be 
implemented in the perturbed Einstein equations
\begin{eqnarray}
\label{t00}
k^{2}\eta -{1\over 2}{\cal H}\dot{h} &=& -{a^2 \delta \rho 
\over 2}  \ ,\\
\label{ti0}
k^{2}\dot{\eta} &=& { a^{2} (p+\rho)\theta \over 2}  \ ,\\
\label{tii}
\ddot{h}+2{\cal H}\dot{h}-2k^{2}\eta &=& -3 a^{2} \delta p \
,\\
\label{tij}
\ddot{h}+6\ddot{\eta}+2{\cal H}(\dot{h}+6\dot{\eta})-2k^{2}\eta &=&
-3 a^{2}(p+\rho)\sigma\ .
\end{eqnarray}
This set of differential equations can be integrated once 
initial conditions on the metric and fluid perturbations are given; 
in this work we adopt adiabatic initial conditions (see \cite{CF}) for the
various components  
and we perform the numerical integration of the system above
for two specific classes of scalar-tensor theories, that will be defined
in the next section.

The numerical integration has been performed by
modifying the standard code CMBFAST \cite{SZ}.
The Q model case was introduced into the
code in \cite{CF} where we investigated the
perturbations behavior in these models.
Here we provide a further extension to cover
EQ models. As a main difference regarding the 
background evolution, the initial conditions 
for the Quintessence have to be searched by 
an iterative method that fixes the initial $\phi$ and 
$\dot{\phi}$ values so that at the present time $a_{0}=1$ 
the Quintessence energy density has the required amplitude. 

\section{Induced gravity and non-minimally coupled scalar field models}

As we mentioned in the Introduction, two subclasses of non-minimally
coupled scalar field theories have been considered 
\cite{Zee}, \cite{Accetta}, \cite{Birrel}. 
Let us define them in the formalism of the previous Section. 

Both these models can be obtained by setting 
\begin{equation}
f(\phi, R)=F(\phi) R\ \ ,\ \ \omega (\phi )=1\ ,
\label{fr}
\end{equation}
so that many of the formulas in the previous section simplify; 
also we take $\beta =1$ requiring that $F$ has the correct physical 
dimensions of $1/G$. Note that all this fixes the link between 
the value of $F$ today and the Newtonian gravitational constant $G$:
\begin{equation}
F_{0}=F(\phi_{0})={1\over 8\pi G}\ .
\label{ftoday}
\end{equation}
Also, this allows to define a time variation of the gravitational 
constant in non-minimally coupled theories, 
\begin{equation}
{G_{t}\over G}=-{F_{t}\over F}\ ,
\label{gtg}
\end{equation}
(where the subscript $t$ indicates differentiation w.r.t. the 
cosmic time $t$) 
that is bounded by local laboratory and solar system experiments 
\cite{Gbounds} to be 
\begin{equation}
{G_{t}\over G}\le 10^{-11}\ {\rm per\ year}\ .
\label{sse}
\end{equation}
There is another independent experimental constraint coming 
from the effects induced on photons trajectories \cite{omega500}. 
As well known, 
by making the transformation $\phi\rightarrow \Phi_{JBD}$ 
such that 
\begin{equation}
{1\over 2}F(\phi )R-{1\over 2}\phi^{;\mu}\phi_{;\mu}
\rightarrow
\Phi_{JBD}R+{\omega_{JBD}\over\Phi_{JBD}}
\Phi_{JBD}^{;\mu}\Phi_{JBD;\mu}\ ,
\end{equation}
the condition $\omega_{JBD}\ge 500$ has to be imposed at the 
present time. 
It is easy to see that in our case this takes the form 
\begin{equation}
\omega_{JBD}={F_{0}\over F_{\phi0}^{2}}\ge 500\ ,
\label{jbd}
\end{equation}
where $F_{\phi 0}$ is the derivative of $F$ w.r.t. $\phi$ 
calculated at the present time. 
As we shall see, this constraint turns out to be the dominant 
one for our models. 

Now let us proceed to the definition of the IG and NMC models. 

In Induced Gravity (IG) models the gravitational constant is directly 
linked to the scalar field itself, as originally proposed in the 
context of the Brans-Dicke theory. We treat here this case by setting 
\begin{equation}
F(\phi )=\xi\phi^{2}\ ,
\label{IG}
\end{equation}
where $\xi$ is the IG coupling constant
In this case equations 
(\ref{ftoday},\ref{sse},\ref{jbd}) become respectively
\begin{equation}
\phi_{0}={1\over\sqrt{\xi 8\pi G}}\ ,
\label{ftodayig}
\end{equation}
\begin{equation}
{\phi_{t0}\over\phi_{0}}\le 10^{-11}\ {\rm per\ year}
\ \ ,
\ \ \xi\le {1\over 2000}\ .
\label{ssejbdig}
\end{equation}
The minimally coupled case is recovered from IG models in the limit 
$\xi\rightarrow 0$; because of equation (\ref{ssejbdig}) this 
implies $\phi_{0}\rightarrow\infty$, and it can be quite easily 
verified that these conditions reduce all the equations 
written in the previous case to ordinary general relativity. 

In non-minimally coupled (NMC) scalar field models the term multiplying 
the curvature scalar $R$ is made of two contributions: the dominant one, 
which is a constant, plus a term  depending on $\phi$; 
minding the constraint on F at the present time, from equation
(\ref{ftoday}), this can be written in the most general way as 
\begin{equation}
F(\phi )\equiv {1\over 8\pi G}+{\widetilde F}(\phi ) - {\tilde F}(\phi_{0})\ .
\label{nmcstructure}
\end{equation}
Then, we choose $\widetilde F$ in equation (\ref{nmcstructure}) as 
\begin{equation}
{\widetilde F}(\phi )=\xi\phi^2\ ,
\label{NMC}
\end{equation}
where again $\xi$ is a coupling constant 
\footnote{Note that we 
define here the coupling constant $\xi$ with the opposite sign w.r.t.  
the standard notation for NMC models.} and the constraints 
(\ref{sse},\ref{jbd}) become 
\begin{equation}
16\pi G\xi\phi_{0}\phi_{t0}\le 10^{-11}\ {\rm per\ year} 
\ \ ,
\ \ 32\pi G\xi^{2}\phi_{0}^{2}\le {1\over 500}\ .
\label{ssejbdnmc}
\end{equation}
Contrary to the IG case, we are now free to set $\phi_{0}$, and the 
ordinary GR case is recovered by taking $\xi\rightarrow 0$. 
Having no restrictions about this point, 
in our numerical integrations we fixed $\phi_{0}=M_P\equiv 
G^{-1/2}$, the Planck mass (in natural units). 
We will only consider here for definiteness the case $\xi>0$. The most 
general case, regarding the background evolution only, 
is discussed in \cite{Chiba}. 

Let us just mention here that one can always map this kind of scalar-tensor 
theories of gravity to canonical general relativity, by means of a conformal 
(Weyl) transformation, 
leading to the so-called Einstein frame (see, e.g. the recent review in
\cite{faraoni}), where the gravity sector of the action takes the standard
Einstein-Hilbert form. In the latter frame, the Quintessence field 
would be minimally coupled with gravity, but it would show 
explicit couplings with all the matter components. 
This mathematical technique is particularly useful 
if one is looking for scaling solutions \cite{Ame}. 
We will not adopt this procedure here, but we will make all our 
calculations in the present physical frame, also called `Jordan frame'.  

Let us elevate now $\phi$ to the role of Quintessence. This  
requires giving it a non-zero potential $V(\phi)$.  
Several potentials have been proposed for the 
Quintessence. In \cite{CDF}, the authors analyzed a cosine 
potential motivated by an ultra-light pseudo Nambu-Goldstone 
boson, while in other works, trying to build a 
phenomenological link to supersymmetry breaking models, 
inverse power potential have been considered \cite{track},\cite{SM}. 
As pointed out in \cite{bine}, inverse power potentials appear 
in supersymmetric QCD theories \cite{sqcd}. 
Here we take the simplest potential of the second class, 
\begin{equation}
V(\phi )={M^5\over\phi}\ ,
\label{v}
\end{equation}
where the mass-scale $M$ is fixed by the level of 
energy contribution today from the Quintessence. 

We are now ready to make some preliminary investigation of the background 
model. We require that the present value of $\Omega_{\phi}$ is $0.6$, 
with Cold Dark Matter at 
$\Omega_{CDM}=0.35$, three families of massless neutrinos, 
baryon content $\Omega_{b}=0.05$ and Hubble constant $H_{0}=50$ Km/sec/Mpc; 
the initial kinetic energy of $\phi$ is not important since it is redshifted 
away during the evolution, so we can fix an equal amount of kinetic and 
potential energy at the initial time $\tau=0$. 

Let us introduce the next Section by fixing 
the compatibility of our models with the 
experimental constraints (\ref{sse},\ref{jbd}). 
A first version of these results, valid only for NMC models, 
can be found in \cite{Chiba}. 

\begin{figure} 
\centerline{
\psfig{file=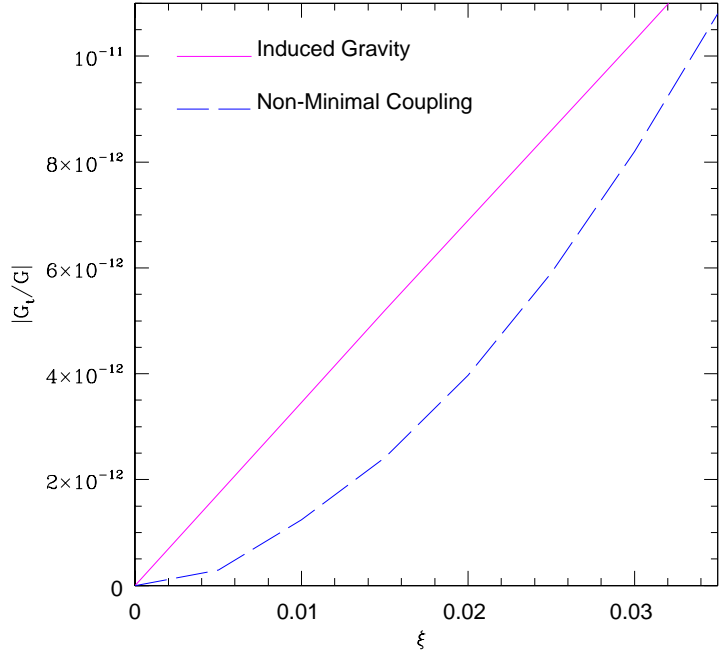,height=5.in,width=5.in}
}
\caption{Numerical analysis of the time variation 
of the gravitational constant versus the QR 
coupling constant in EQ models.}
\label{fsse}
\end{figure}

First, we integrate equations 
(\ref{Friedmann},\ref{KG}) to compare  
with the experimental constraint of Eq.(\ref{sse}). 
The results are shown in Fig.\ref{fsse}, where 
$|G_{t}/G|$ at the present time is shown as a function of 
$\xi$. Both for NMC and IG, the limit roughly is 
\begin{equation}
\xi\lesssim 3\times 10^{-2}\ .
\label{consse}
\end{equation}
However, as we anticipated, the stronger constraint comes 
from Eq.(\ref{jbd}); it is simple to see that in our models 
Eqs.(\ref{ssejbdig},\ref{ssejbdnmc}) become 
\begin{equation}
\xi\lesssim 5\times 10^{-4}\ \ {\rm IG\ case}\ ,
\label{conjbdig}
\end{equation}
\begin{equation}
\xi\lesssim 5\times 10^{-3}(\sqrt{G}\phi_{0})^{-1}
\ \ {\rm NMC\ case}\ .
\label{conjbdnmc}
\end{equation}
In the next section we will explore the effects on the 
cosmological perturbations spectra of EQ models, 
also considering values of $\xi$ 
beyond the above constraints, 
in order to better illustrate its effect on 
the cosmological equations. 
Then, we will discuss how future CMB experiments like MAP and 
Planck will be able to detect features of the present models 
within the range allowed from Eqs.(\ref{conjbdig},\ref{conjbdnmc}). 

\section{QR-effects on cosmological perturbations}

Here we present the results coming from the integration  
of the complete set of equations of Sec. II. The numerical
integration of this set of equations has not been performed 
before, and we obtain several new and interesting effects
concerning cosmologies with a coupling between Quintessence and the 
Ricci curvature scalar $R$, that we name `QR-effects'; 
we discuss them in the following subsections. 

Let us now set initial conditions for the 
perturbation equations, referring to \cite{CF} for 
an extensive treatment. We adopt 
isoenthropic (i.e. adiabatic) initial 
conditions; in the minimal coupling case they are quite 
simple: everything is initially zero except for the 
metric perturbation $\eta$. It is easy to check that 
these conditions remain valid also in the present case. 
In fact, adiabaticity is imposed on each fluid separately, 
by requiring that the entropy perturbations is equal 
to zero initially for each pair of fluid components, 
including Quintessence \cite{CF}; these conditions 
do not depend on the coupling of a given component with $R$. 

As we anticipated, the scalar-tensor theories of gravity that we 
consider leave several characteristic imprints on cosmological 
perturbations spectra. Also, both IG and NMC models, although 
for different coupling constant ranges, 
show a remarkably similar behavior. 
For clearness, we shall 
treat first the features related to the 
background evolution and successively the 
genuine QR-effects on perturbations.

\subsection{QR-effects on the background: enhanced Hubble length 
growth and $\Omega_{matter}>1$}

Let us consider the Hubble length first. 
The integration of Eqs.(\ref{Friedmann},\ref{Friedmann2}) 
with the potential (\ref{v}) shows that 
the time derivative of the Hubble length, 
$H^{-1}_{t}(z)$, {\it increases} at non-zero redshifts compared 
with the ordinary Quintessence case, both for NMC and IG models. 
Therefore, fixing the Hubble length 
at present as we do, implies that in the past 
it was smaller than in minimally coupled models. 
This effect is clearly displayed 
by Fig.\ref{fhubble}, where the comoving 
Hubble length as a function of $z$ is shown 
(for simplicity we plot the IG case only, the NMC one
being completely equivalent). This feature has been 
already noted in the context of pure Brans-Dicke 
theories \cite{liddle}. 
The sharp change in the time dependence of $H^{-1}$  
at small redshifts is due to the Q-field, that 
dominates the cosmological evolution at later times. 

The source of the enhanced Hubble length growth in our  
models is the last term in the 
Einstein equation (\ref{Friedmann2}); as we will 
show in a moment, this term is quite large and positive, 
being also responsible for most of the features that 
we shall see later concerning the cosmological 
perturbation spectra. 

A related interesting point is that our model predicts 
a small change in $H$ which mimics a change in 
the number of massless neutrinos at the 
Nucleosynthesis epoch (see \cite{SARKAR} for an extensive 
overview). At this time Quintessence is very 
subdominant and the cosmological evolution is governed by the 
equation 
\begin{equation}
H^{2}\simeq {\rho_{fluid}\over 3F(\phi )}\ ;
\label{nucleo1}
\end{equation}
since in our models $F(\phi )<F(\phi_{0})$ at any past time, 
the shift in the value of $H^{2}$ due the time variation 
of the gravitational constant in EQ models is given by: 
\begin{equation}
{\Delta H^{2}\over H^{2}}=1-{F(\phi )\over F(\phi_{0})}\ . 
\label{dh2}
\end{equation}
As a function the shift $\Delta N$ of the number of 
relativistic species at Nucleosynthesis, the above quantity may 
be written as 
\begin{equation}
1-{F(\phi )\over F(\phi_{0})}=
{7\Delta N /4\over 10.75+7\Delta N/4}\ .
\label{dn}
\end{equation}
Therefore, the shift $\Delta N^{QR}$ predicted in our models 
is 
\begin{equation}
\Delta N^{QR}=-6.14\times {F(\phi_{0})-F(\phi )\over 
F(\phi_{0})-2F(\phi )}\ .
\label{dnqr}
\end{equation}
It is worthwhile to note that for models 
satisfying Eq.(\ref{jbd}), the predicted $\Delta N^{QR}$ is 
at the level of $10\%$, thus being well below the current 
experimental constraints from the Nucleosynthesis. 

\begin{figure}
\centerline{
\psfig{file=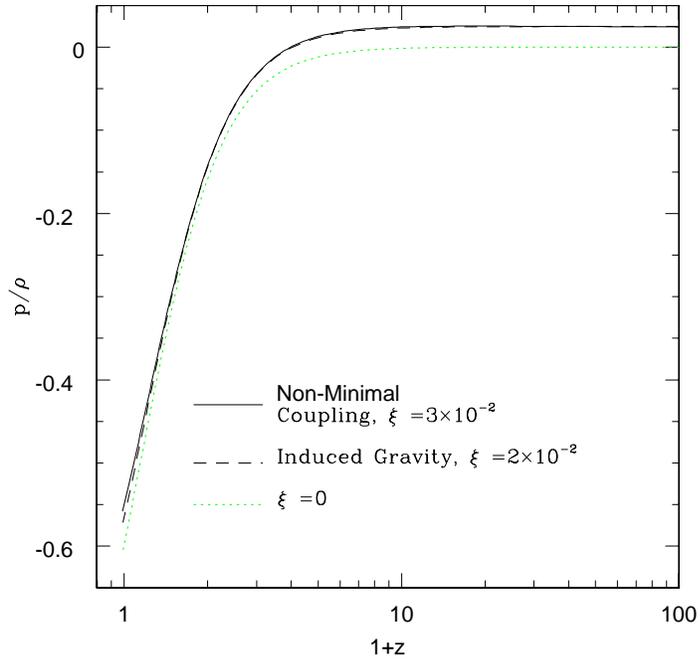,height=5.in,width=5.in}
}
\caption{Time behavior of the cosmic equation of state in EQ models 
versus ordinary Quintessence.}
\label{feos}
\end{figure}

Let us consider now the effects of our scenario on the 
cosmological equation of state. The ${\cal H}\dot{F}/F$ 
term appears also in the effective fluid pressure in 
Eq.(\ref{rhop}), causing the following interesting feature 
in the behavior of the equation of state, shown in 
Fig.\ref{feos}. As it is evident, in the matter dominated 
era $p/\rho >0$ up to $1+z\approx 5$, when the Quintessence starts 
to dominate. Thereafter, the cosmic expansion starts to 
accelerate because of the vacuum energy stored in the Quintessence 
potential. Thus we have the apparent paradox that in the 
matter dominated era the total pressure is non-zero and 
positive: this 
is not surprising since it can be brought back to the dynamics 
of the scalar field itself in scalar-tensor theories of gravity. 
Corresponding to its positive value in the matter dominated era, 
the equation of state at present, when Quintessence dominates, 
is slightly above its value for Q models. 
In other words, we found that the Quintessence contribution to 
the equation of state in our models, $p_{\phi}/\rho_{\phi}$, 
does not change significantly in our case with respect to Q 
models; we found indeed 
\begin{equation}
-1\lesssim {p_{\phi}\over\rho_{\phi}}\lesssim -0.9
\label{eosphi}
\end{equation}
for all the cases considered. This is 
well within the range of values for which the Quintessence 
is mimicking a cosmological constant 
\cite{eqofstate}, \cite{garna}. 

Let us now come to the $\Omega_{matter}>1$ effect. 
This interesting and very peculiar occurrence can be 
understood by looking at the behavior of the various components 
of the energy density in Eq.(\ref{Friedmann}) and is obviously 
connected with the effect on the equation of state just described. 
After dividing both members by ${\cal H}^{2}$, 
the Friedmann equation takes the form 
\begin{equation}
1=\Omega (z)_{matter}+\Omega (z)_{radiation}+\Omega (z)_{\phi}\ ,
\label{1omega}
\end{equation}
where it must be noted that $\Omega_{\phi}$ is actually made  
of three terms, namely  
\begin{equation}
\Omega (z)_{\phi} = 
\Omega (z)_{\phi}^{K}+\Omega (z)_{\phi}^{P}+\Omega (z)_{\phi}^{QR}\ .
\label{phiomega}
\end{equation}
While $\Omega_{\phi}^{K}$ and $\Omega_{\phi}^{P}$ are the 
generalization of the kinetic and potential energy densities 
in scalar-tensor theories, the really new component is 
\begin{equation}
\Omega_{\phi}^{QR}=-{F_{\phi}\dot{\phi}\over F{\cal H}}\ , 
\label{omegaqr}
\end{equation}
which, as we already noted, is {\it negative} if 
$\dot{\phi}>0$. Its amplitude is fixed essentially 
by the dynamics of the scalar field; as we anticipated, 
this term turns out to be important for the background 
evolution. The reason is the following. 
In all the cases considered, the scalar field evolution is
slow, so that $\ddot{\phi}$ and the time variation of the 
potential in the Klein-Gordon equation can be neglected. 
Let us consider the radiation dominated era for simplicity: 
$a=\dot{a}_{rad}\tau$, where $\dot{a}_{rad}$ is a constant. 
Therefore, it is immediate to check that the approximate solution 
of the Klein Gordon equation is 
\begin{equation}
\phi =\phi_{initial}-{\dot{a}^{2}_{rad}V_{\phi}\over 20}
(\tau^{4}-\tau^{4}_{initial})\ .
\label{kgapprox}
\end{equation}
In the ideal case where the scalar field evolves for a large time 
so that only the term proportional to $\tau^{4}$ is important, 
we see that $\dot{\phi}/\phi\propto 1/\tau\propto {\cal H}$; 
in this case the term in Eq.(\ref{omegaqr}) would be of order 
unity. In the real case these arguments are 
weakened since the scalar field does not have a perfect 
slow-rolling dynamics, and it does not evolve enough to 
become much larger than its initial value; nevertheless 
this qualitatively explains why we found 
$\Omega_{\phi}^{QR}\sim 10^{-2}$ 
for models satisfying the constraints 
(\ref{jbd}), and for a time interval roughly covering 
all the post-equality cosmological history. 

\begin{figure}
\centerline{
\psfig{file=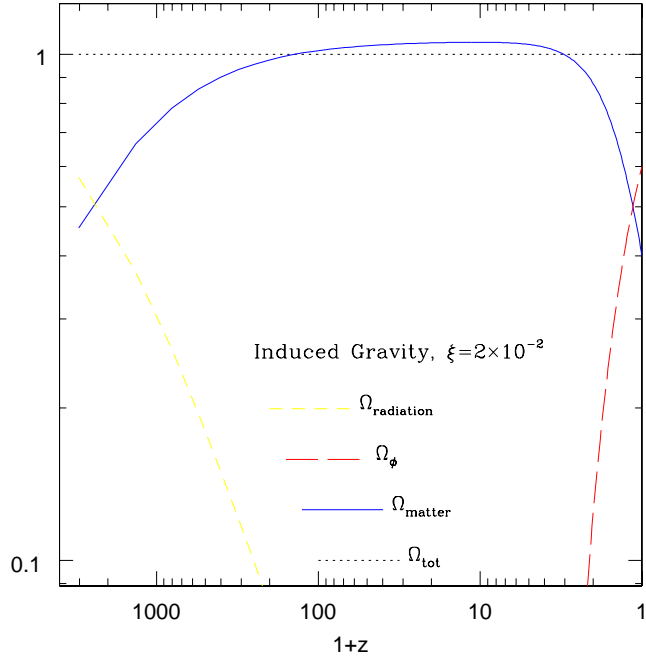,height=5.in,width=5.in}
}
\caption{Time behavior of the $\Omega$ parameters relative 
to matter, radiation and Quintessence.}
\label{fomega}
\end{figure}

\begin{figure}
\centerline{
\psfig{file=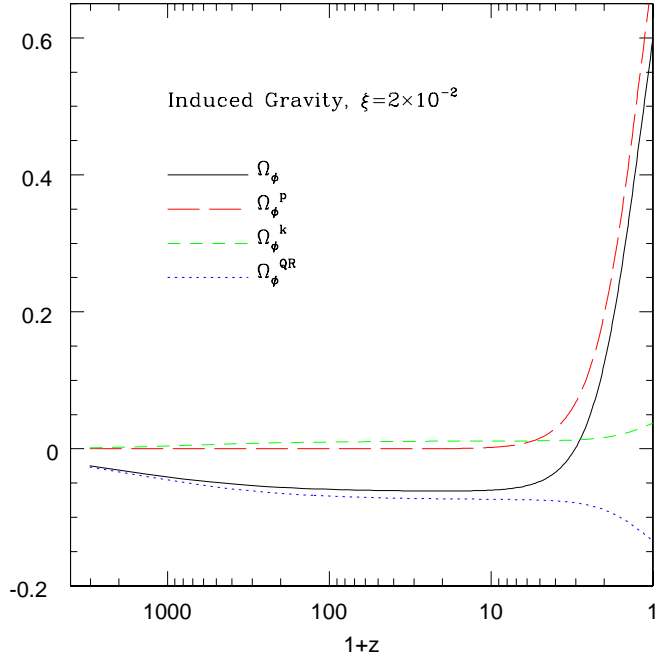,height=5.in,width=5.in}
}
\caption{Time behavior of the $\Omega_{\phi}$ parameters relative 
to the potential, kinetic and purely QR terms.}
\label{fomegaphi}
\end{figure}

Fig.\ref{fomega} shows the various contributions 
to the cosmic density parameters as a function of redshift. 
The matter radiation equality epoch is clearly visible, as well 
as the matter dominated era, and, finally, the Quintessence dominated 
era at very small redshifts. Also, the sum (identically equal to 1) 
is shown, and it is immediately seen that in the matter dominated era 
one has 
\begin{equation}
\Omega_{matter}>1\ .
\label{1ltomegamatter}
\end{equation}
As we already anticipated this is only an apparent paradox, 
because of the presence of the {\it negative} energy component 
in the Einstein equation (\ref{Friedmann}), explicit in 
Eq.(\ref{omegaqr}). Figure \ref{fomegaphi} shows 
the various contributions to the Quintessence  
energy density. As it can be seen, for the chosen value of the coupling 
constant $\xi$, $\Omega_{\phi}^{QR}$ reaches 
values of a few percent and is responsible for the 
condition (\ref{1ltomegamatter}). 

This completes a rapid survey of the 
features regarding the cosmological background evolution; 
some of them have a relevant influence on the perturbation 
behavior, that is the subject of the next subsection. 

\subsection{QR-effects on the CMB: Integrated Sachs-Wolfe effect, 
horizon crossing delay and reduced acoustic peaks} 

The phenomenology of CMB anisotropies in EQ models 
is rich and possesses distinctive features. 

In the top left panel of Fig.\ref{fig}, 
the effect of increasing $\xi$ on 
the power spectrum of COBE-normalized CMB anisotropies 
is shown. Note that we plotted cases also exceeding the limit 
(\ref{jbd}), to make clearer the perturbations behavior 
in EQ scenarios. 
The rise of $\xi$ makes substantially three 
effects: the low $\ell$'s region is enhanced, the oscillating 
one attenuated, and the location of the peaks shifted 
to higher multipoles. Let us now explain these effects. 
The first one is due to the integrated Sachs-Wolfe effect, 
arising from the change from matter to Quintessence dominated 
era occurred at low redshifts. This occurs also in ordinary Q models, but 
in EQ this effect is enhanced. 
Indeed, in ordinary Q models the dynamics of $\phi$ is governed by 
its potential; in the present model, one more independent 
dynamical source is the coupling between the Q-field and the Ricci
curvature $R$. As it can be easily understood by the 
Lagrangian in equation (\ref{action}), the scalar field $\phi$ 
evolves as dictated by the effective potential  
\begin{equation}
V_{eff}(\phi )=V(\phi )-{1\over 2}F(\phi )R\ .
\label{veff}
\end{equation}
As it is clear from equation (\ref{riccih}), 
$R$ is positive in the matter dominated era, 
($a(t)\sim t^{2/3}$). Thus, from (\ref{veff}), 
after differentiating with respect to $\phi$, both the forces coming from  
$V_{eff}$ are {\it negative}, pushing together
the field $\phi$ towards increasing values.
In conclusion, the dynamics of 
$\phi$ is boosted by $R$ together with its potential $V$. 
As a consequence, part of the COBE normalization at 
$\ell=10$ is due to the Integrated Sachs-Wolfe effect; 
thus the actual amplitude of the underlying 
scale-invariant perturbation spectrum 
gets reduced. This is the main reason why the oscillating part 
of the spectrum, both for polarization and temperature, is below 
the corresponding one in Q-models. 

\begin{figure}
\centerline{
\psfig{file=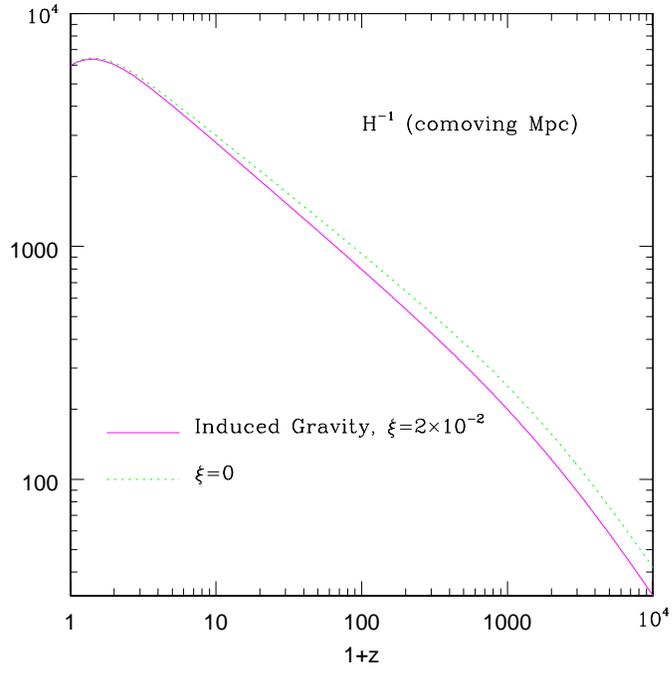,height=5.in,width=5.in}
}
\caption{Time behavior of the Hubble length in EQ models 
versus ordinary Quintessence.}
\label{fhubble}
\end{figure}

\begin{figure}
\centerline{
\psfig{file=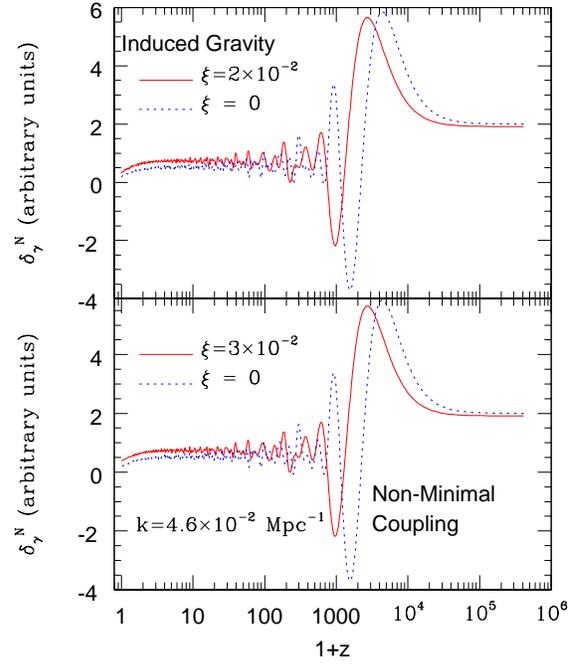,height=5.in,width=5.in}
}
\caption{Comparison of the time behavior of the photon density
fluctuations for the scale shown for EQ and Q models.}
\label{fdeltag}
\end{figure}

There is however another effect that slightly reduces the 
amplitude of the acoustic oscillations. We have seen in 
Fig.\ref{fhubble} that the Hubble length was smaller in the past 
in EQ than in Q models. This has the immediate 
consequence that the horizon crossing of a given 
cosmological scale is delayed. This 
is manifest in Fig.\ref{fdeltag}, where we have 
plotted the photon density perturbation in 
the Newtonian gauge $\delta_{\gamma}^{N}$; we choose this quantity 
since it is simply 4 times the dominant term of the CMB temperature 
fluctuations \cite{HSWZ}. Its expression in terms of the quantities 
in the synchronous gauge is 
\begin{equation}
\delta_{\gamma}^{N}=\delta_{\gamma}+
{\dot{h}+6\dot{\eta}\over 2k^{2}}{\dot{\rho}\over\rho}\ . 
\label{deltagn}
\end{equation}
The scale shown in Fig.\ref{fdeltag} 
is chosen so that it reenters the horizon 
roughly between matter-radiation equality and 
decoupling. Both in the IG and 
NMC cases, it is evident that the 
oscillations start later than 
in ordinary Q models. As well known, 
the amplitude of the 
acoustic oscillations slightly decreases if the 
matter content of the universe at decoupling is 
increased \cite{CF,CDF}. 

Finally, note how the location of the acoustic peaks 
in term of the multipole $\ell$ at which the oscillation occurs, is shifted 
to the right. Again, the reason is the time dependence 
of the Hubble length, which at decoupling, 
subtended a smaller angle on the sky. It is straightforward 
to check that the ratio of the peak multipoles in 
Fig.\ref{fig} coincides numerically with the 
the ratio of the values of the Hubble lengths at decoupling 
in Fig.\ref{fhubble} in EQ and Q models. 

These considerations do not change at all for NMC models. 
Really, IG and NMC models present, for different values 
of $\xi$, remarkably similar features, yielding 
a genuine signature of scalar tensor-theories in the 
cosmological perturbations spectra. 

\begin{figure}
\centerline{
\psfig{file=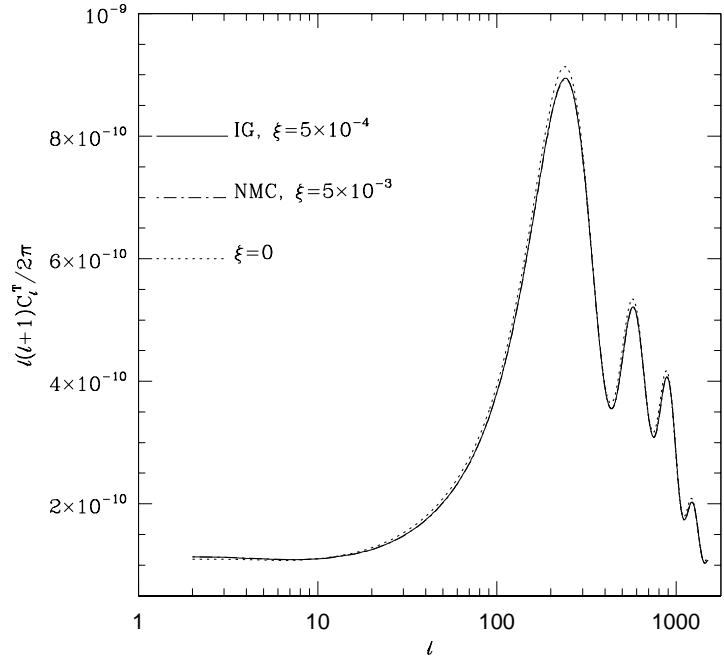,height=5.in,width=5.in}
}
\caption{CMB temperature Perturbations for IG and NMC models for $\xi$ 
satisfying the constraints from solar system experiments.}
\label{freal}
\end{figure}

\begin{figure}
\centerline{
\psfig{file=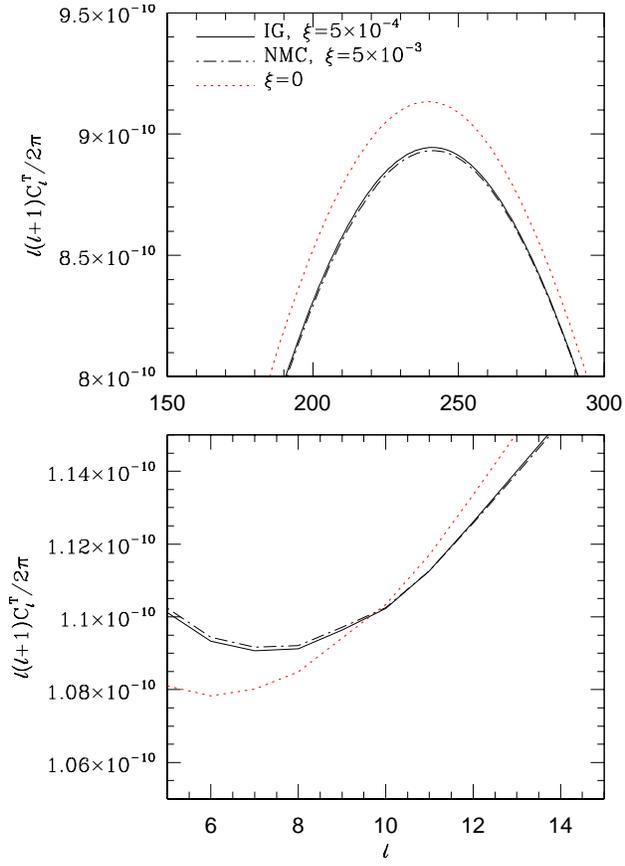,height=5.in,width=5.in}
}
\caption{CMB temperature perturbations for IG and NMC models, 
for $\xi$ satisfying the constraints from solar system experiments: 
first acoustic peak and low $\ell$s power zoomed.}
\label{frealzoom}
\end{figure}

Let us consider now realistic cases respecting the constraints 
from Eq.(\ref{jbd}). Figs.\ref{freal},\ref{frealzoom} 
show the temperature perturbation spectra for NMC and IG 
cases with the indicated coupling constants. The 
effects described previously are evident particularly 
in Fig.\ref{frealzoom}, where the changes in the 
first acoustic peak (top) and in the power at low $\ell$'s 
(bottom) have been zoomed; also, the slight difference 
between IG and NMC models is visible. 
We notice that features of this amplitude in the 
CMB spectra, induced by models satisfying the 
existing constraints from Eqs.(\ref{sse},\ref{jbd}) 
are detectable by the future 
generation of CMB experiments; in particular, MAP 
and Planck will bring the accuracy on the CMB power 
at percent level up to $\ell\simeq 1000$ \cite{CMBFUTURE}. 

\subsection{QR-effects on matter perturbations: 
power-spectrum decrease and peak shift}

\begin{figure}
\centerline{
\psfig{file=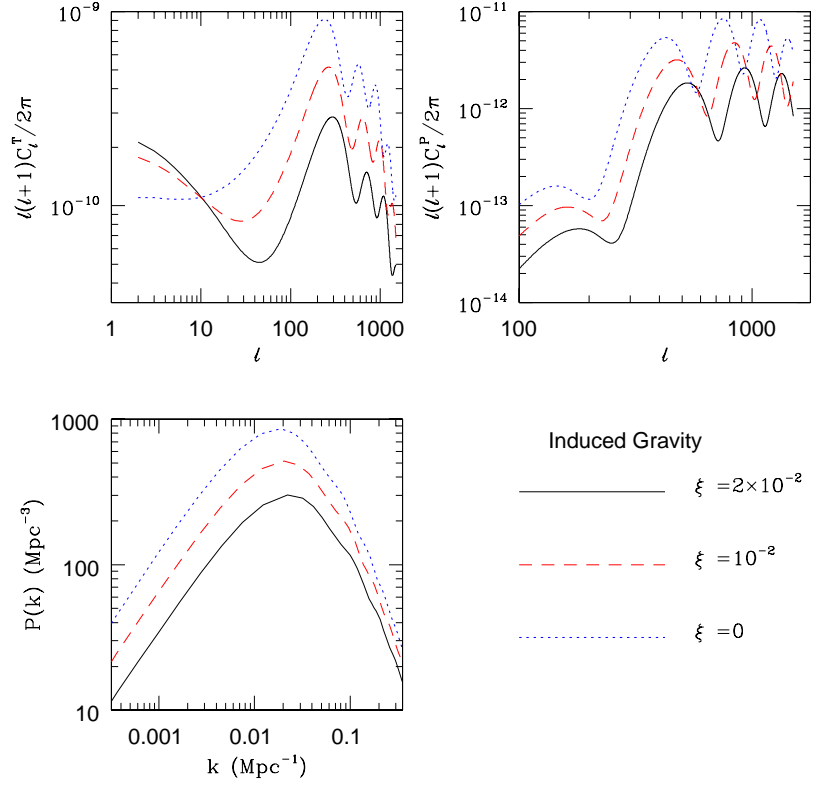,height=5.in,width=5.in}
}
\caption{Perturbations for IG models for various 
values of $\xi$: 
CMB temperature (top left), polarization (top right), and 
matter power spectrum (bottom).}
\label{fig}
\end{figure}

After decoupling, the different models considered 
in Fig.\ref{fig} evolve until the present, when we 
snapshot the matter power-spectrum in the bottom left panel. 

Soon after their introduction, Q-models were 
considered more appealing than those involving a cosmological 
constant term because of their 
capability to shift the power-spectrum toward larger scales without 
increasing its overall amplitude, which would have 
required an antibias mechanism. 
We find here that this effect is enhanced if a QR-coupling 
exists. This is evident in both the bottom right panel in 
Fig.\ref{fig}. The spectra are COBE-normalized 
as it is evident in the top panel. 
For increasing $\xi$, the spectra loose power. 
The reason of this behavior is that the CMB spectra 
include different effects together with the true 
perturbation amplitude; on the large scales 
measured by COBE, the matter perturbations add 
with the large Integrated Sachs-Wolfe effect; 
the greater is $\xi$, the stronger 
being the Integrated Sachs-Wolfe effect, 
the weaker the true perturbations amplitude, 
as we pointed out in the previous subsection. 
This causes the power-spectrum decrease that is 
well visible in the figure. 

The other effect is the slight shift of the location 
of the peaks toward larger wavenumbers. Again, this is due to the time 
dependence of $H^{-1}$; since it is smaller in extended Quintessence
models 
than in ordinary Quintessence ones, the horizon crossing 
is delayed for all the cosmological scales, for the given value 
of $H_{0}$. 

\begin{figure}
\centerline{
\psfig{file=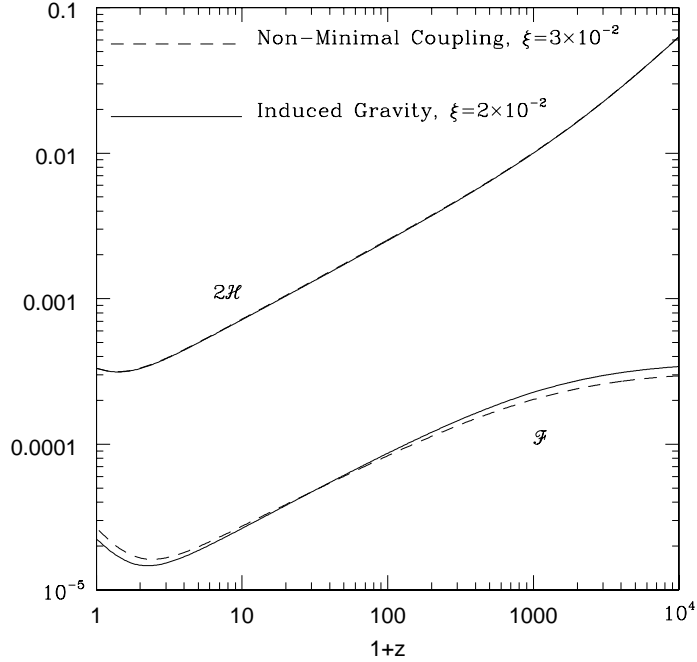,height=5.in,width=5.in}
}
\caption{Time behavior of the friction term (in arbitrary units) 
arising from the shear perturbation 
in EQ models compared with its 
cosmological counterpart.}
\label{ffriction}
\end{figure}

These are the most prominent features concerning the power-spectrum. 
In principle however, there are terms in the cosmological 
perturbation equations that could make some relevant effects. 
We search them as terms that do not multiply fluctuations 
in the scalar field, since the latter are negligible from 
the point of view of structure formation \cite{CF}. 
Looking indeed at Eq.(\ref{shear}), the last 
term in the r.h.s. could play some role: it is 
the shear perturbation associated with the Quintessence and it 
should be noted that it is not present in ordinary Q models. 
Looking at Eq.(\ref{tij}), it is immediate to verify that 
this term produces a sort of excess friction in the dynamics 
of the quantity $\dot{h}+6\dot{\eta}$ in addition to the 
cosmological Hubble drag term $2{\cal H}$ 
in the l.h.s.: we define it as 
\begin{equation}
{\cal F}={\dot{F}\over F}\ .
\label{termshear}
\end{equation}
Its relevance compared to ${\cal H}$ has been already 
discussed when we dealt with the $\Omega_{\phi}^{QR}$ quantity 
of Eq.(\ref{omegaqr}). 
As it is evident in Fig.\ref{ffriction}, ${\cal F}$ 
is not so important during the evolution since it is only a few  
percent of the Hubble drag during all the evolution. 
Although ${\cal F}$ clearly plays the role of a sort of integrated shear 
effect, it is less important than those described at the beginning of this 
subsection. 

These effects change the matter power-spectrum today in a way 
that we will better explore in a future work. Here 
we make a first comparison with the known expectations 
concerning the spectrum normalization at $8h^{-1}$ Mpc, 
$\sigma_{8}$. 
Recently the cluster abundance in Q models has been analyzed 
\cite{sigma8}. An empirical formula for $\sigma_{8}$ in these 
models has been found as 
\begin{equation}
\sigma_{8}=(0.5-0.1\Theta )\Omega_{m}^{-\gamma (\Omega_{m},\Theta )}\ ,
\label{sigma8emp}
\end{equation}
where 
\begin{equation}
\Theta =(n-1)+(h-0.65)\ \ ,\ \ \gamma (\Omega_{m},\Theta )=
0.21-0.22{p_{\phi}\over\rho_{\phi}}+0.33\Omega_{m}+0.25\Theta\ ;
\end{equation}
$n$ is the spectral index (1 in our scale-invariant case), 
$h$ is the present Hubble constant in units of 
100km$\,$s$^{-1}\,$Mpc$^{-1}$ and $\Omega_{m}$ the 
matter energy amount today. 
The existing experimental constraints (see \cite{sigma8}) 
may be expressed as follows:  
\begin{equation}
\label{consigma8}
\sigma_{8}\Omega_{m}^{\gamma}=0.5\pm 0.1\ .
\end{equation}
Our scenario 
is not significantly constrained by Eq.(\ref{consigma8}). 
For the models shown in Fig.\ref{fig}, we found 
\begin{eqnarray}
\sigma_{8}&=&0.525\ \ {\rm for}\ \xi=2\times 10^{-2}\nonumber\\
\sigma_{8}&=&0.623\ \ {\rm for}\ \xi=10^{-2}
\label{sigma8num}\\
\sigma_{8}&=&0.725\ \ {\rm for\ ordinary\ Q\ models.}
\end{eqnarray}
It is easy to verify that the constraint in Eq.(\ref{consigma8}) 
is satisfied for $\xi\lesssim 10^{-2}$; the same limit 
for NMC models is $\xi\lesssim 2\times 10^{-2}$. 
It is remarkable however that future experiments 
will be able to provide much more accurate measurements 
of the matter power spectrum \cite{LSSFUTURE}.

\section{Summary and concluding remarks}

Our work is based on the possibility that the cosmological vacuum energy 
that seems required  to explain the data from high-redshift type Ia
Supernovae resides in the potential energy of 
a slowly rolling scalar field or Quintessence. We considered 
models in which the Quintessence scalar field is non-minimally coupled
with the Ricci curvature scalar $R$, that we named Extended 
Quintessence. 

With this aim, and based on a technique obtained in some recent works 
\cite{HW1,HW2,CF}, we integrated  
the full linear cosmological perturbation equations for 
generalized Einstein gravity theories. In this framework we 
investigated the effects produced by two distinct 
Extended Quintessence models, in which 
the gravitational part of the Lagrangian is 
\begin{equation}
{1\over 16\pi G}R\rightarrow {F(\phi )\over 2}R
\ \ \ {\rm with}\ 
\nonumber
\end{equation}
\begin{equation}
F(\phi )= \xi\phi^{2}\ {\rm (IG\ models)\ \ \ and}\ \ \  
F(\phi )= {1\over 16\pi G}+\xi (\phi^{2}-\phi_{0}^{2})
\ {\rm (NMC\ models)}\ ,
\label{signmc}
\end{equation}
$\phi_{0}$ indicating the Q-value today. 

Quintessence models are characterized by a potential energy 
that is comparable to the matter energy density today. 
We choose the simplest inverse power potential 
\begin{equation}
V(\phi )={M^{5}\over\phi}\ ,
\label{sv}
\end{equation}
with the constant $M$ fixed by requiring that 
the Quintessence energy density today yields
$\Omega_{\phi}=0.6$. 

The first check we made by integrating our equations, was whether our
results are compatible with the bounds from the solar system experiments: 
we found that these constraints are satisfied  
if $\xi\lesssim 5 \times 10^{-4}$, for IG, and $\xi \lesssim 5 \times
10^{-3} (\sqrt{G} \phi_0)^{-1}$, for NMC models. 
We went then to a more detailed analysis 
of the effects on the power-spectra obtained, that 
we named QR-effects. We found several features that could help in 
discriminating these models from ordinary Quintessence. 

In particular, the Integrated Sachs-Wolfe effect, 
caused by the time variation of the gravitational 
potential between last scattering and the present time, which is 
already active in ordinary Q-models, 
is now enhanced.
This can be understood by considering 
the Klein-Gordon equation governing the time evolution 
of $\phi$. It is easily seen that the coupling 
with $R$ induces a new source of effective potential 
energy; the latter is ineffective in the 
radiation dominated era, when $R \approx 0$, but becomes 
important during matter and scalar field dominance, when 
it originates the effective potential 
\begin{equation}
V_{QR}=-{1\over 2}F(\phi )R\ .
\label{sveff}
\end{equation}
It is therefore immediate to realize that the force $dV_{QR}/d\phi$ 
in the Klein Gordon equation simply 
adds to the one coming from the true 
potential $dV/d\phi$ from (\ref{sv}), having 
the same sign and therefore enhancing the Integrated Sachs-Wolfe effect. 
As a consequence, part of the COBE normalization 
is now due to the latter effect and the cosmological perturbation 
amplitude, including also the oscillating region of 
the CMB spectrum, is reduced; this is 
evident in the CMB polarization and temperature patterns, as well as 
in the matter power-spectrum today. 
Moreover, the acoustic peaks and the power-spectrum turnover are 
displaced to smaller scales; the reason being that the Hubble length 
$H^{-1}$ grows more rapidly in these theories 
than in ordinary Q-models, delaying - for a fixed value of $H_0$ - the
horizon crossing of 
any scale larger than the Hubble radius at the
matter-radiation equality, and slightly decreasing 
the amplitude of the acoustic oscillations. 

Another independent QR-effect comes from the change of 
the fluid shear $\sigma$ arising in generalized Einstein 
theories. From the Einstein equations it turns out that 
the new terms in $\sigma$ induce an additional friction 
to the growth of the gauge-invariant gravitational potential 
$\Psi$, besides that due to the Hubble drag. This makes 
the growth of $\Psi$ weaker, and, since in adiabatic models 
the acoustic oscillations are essentially driven by this quantity, 
this results in a reduced amplitude for the acoustic peaks. 

For what concerns large-scale structure formation, 
we also considered the effect of the extra term 
in the fluid shear arising from the QR-coupling. 
It produces a sort of friction in the dynamics of the 
metric perturbations, in addition to the 
genuine cosmological friction. Although interesting, 
we found that this effect is negligible compared to the effect due  
to the Integrated Sachs-Wolfe effect that changes the normalization 
to COBE data. 

It is also remarkable that similar features 
occur both in IG and NMC models, suggesting the existence 
of an extended Quintessence phenomenology that is the signature of a large 
class of scalar-tensor theories in the cosmological perturbations. 

This is a brief summary of the results we obtained 
in this class of Extended Quintessence models. Of course, this work does 
not answer all the questions nor it explores all the aspects, but the results 
we obtained show distinctive and promising features at the point 
that we believe it should be seriously taken into account, 
especially in favor of the hints on the existence of scalar 
fields and on their possible couplings with $R$ coming 
from fundamental theories. 
An important problem to face is which 
effects are caused by the fact that we require that the 
field coupled with $R$ is a Quintessence, and which 
instead come from the scalar-tensor theories themselves. The enhanced 
Integrated Sachs-Wolfe effect appears to be mostly determined by the
extra effective potential coming from the non-minimal coupling; 
on the contrary, the effects at decoupling appear to be caused 
mostly by the true scalar field potential, 
since at that time the Ricci scalar $R$ is much smaller than 
it is now. However, all these considerations, together for example 
with the exploration of other scalar field potentials and
more general gravitational sectors in the Lagrangian, would deserve a
separate work. 

The results obtained here are potentially testable by the upcoming 
experiments which aim at gaining detailed information on   
cosmological parameters, both from the CMB \cite{CMBFUTURE} and 
from the large-scale structure \cite{LSSFUTURE}. 

\acknowledgements

We warmly thank Luca Amendola, Robert Caldwell and 
Karsten Jedamzik for precious comments and useful discussions. 

While completing this paper a preprint by Chen and Kamionkowski 
\cite{kamion} has appeared in which the CMB temperature and 
polarization patterns produced by a pure JBD field in a 
standard Cold Dark Matter cosmology have been considered. 
Although there is no overlap with our Quintessence 
field, it is worthwhile to note that, for what concerns 
the acoustic peak locations, their results show a similar 
dependence on the $\omega_{JBD}\propto 1/\xi$ parameter. 

\begin{thebibliography}{}
\bibitem{Stain1} 
P.J. Peebles, B. Ratra, Astrophys. J. 352, L17 (1988); 
R.R. Caldwell, R. Dave, P.J. Steinhardt, 
Phys. Rev. Lett. 80, 1582 (1998). 
\bibitem{Stain2} P.G. Ferreira, 
M. Joyce , Phys. Rev. D 58 023503 (1998); 
P.T.P. Viana, A.R. Liddle,  Phys. Rev. D57, 674 (1998).
\bibitem{CDF} K. Coble, S. Dodelson, J. Friemann, Phys. Rev. D 55,
1851 (1997).
\bibitem{Ratra} B. Ratra, P.J. Peebles, Phys. Rev. D 37, 3406
(1988).
\bibitem{CF} F. Perrotta, C. Baccigalupi, Phys. Rev. D 59, 123508
(1999).
\bibitem{Perlm} S. Perlmutter {\it et al.}, Nature 391, 51 (1998); 
A.G. Riess {\it et al.}, Astron. J. 116, 1009 (1998). 
\bibitem{garna} P.M. Garnavich {\it et al.}, 
ApJ in press, preprint astro-ph/9806396. 
\bibitem{Lambda} A.D. Dolgov, {\it Lecture presented 
at the 4$^{th}$ Colloque Cosmologie}, Paris (1997).
\bibitem{Sahni} V. Sahni, A. Starobinski, 
preprint astro-ph/9904398 (1999).
\bibitem{track}
P.J. Steinhardt, L. Wang, I. Zlatev, 
Phys. Rev. D 59,  123504 (1999);
I. Zlatev, L. Wang,  P.J. Steinhardt, Phys. Rev. Lett. 82, 896 (1999);
A.R. Liddle, R.J. Scherrer, Phys. Rev. D 59, 023509 (1999).  
\bibitem{Roll} P.G. Roll, R. Krotkov, R.H. Dicke, Ann. Phys. (N.Y.) 26, 
442 (1964).
\bibitem{carroll} S.M. Carroll, Phys. Rev. Lett. 81, 15, 3097 (1998).
\bibitem{ame0} L. Amendola, preprint astro-ph/9908023 (1999). 
\bibitem{Chiba} T. Chiba, to appear in Phys. Rev. D, 
preprint gr-qc/9903094 (1999).
\bibitem{Ame} L. Amendola, Phys.Rev. D60 043501 (1999), 
astro-ph/9904120.
\bibitem{Uzan} J.P. Uzan, Phys. Rev. D 59, 123510 (1999).
\bibitem{exte} D. La, P.J. Steinhardt, Phys. Rev. Lett. 62, 
376 (1989); F.S. Accetta, J.J. Trester, Phys. Rev. D 39,  2854 (1989); 
E.J. Weinberg, Phys. Rev. D 40, 3950 (1989). 
\bibitem{JBD} P. Jordan, Z. Phys. 157, 112 (1959); 
C. Brans, R.H. Dicke, Phys. Rev. 124, 925 (1959). 
\bibitem{dolgov} A.D. Dolgov, In {\it The Very Early Universe}, 
eds. G.W. Gibbons, S.W. Hawking, S.T.C. Siklos, 
Cambridge University Press, Cambridge, p. 449 (1983). 
\bibitem{Zee} F. Zee, Phys. Rev. Lett. 42, 417 (1979).
\bibitem{Accetta} B.J. Spokoiny, Phys. Lett. B 147, 39 (1984); 
F.S. Accetta, D.J. Zoller, M.S. Turner, Phys. Rev. 
D 31, 3064 (1985); M.D. Pollock, Phys. Lett. B 156, 301 (1985);
F. Lucchin, S. Matarrese, M.D. Pollock, Phys. Lett. B 167, 
163 (1986). 
\bibitem{Bellido} J. Garcia-Bellido, Phys. Rev. D 55, 4603 (1997).  
\bibitem{Birrel} N.D. Birrel, P.C. Davies, Quantum Fields in Curved
Spaces, Cambridge Univ. Press, Cambridge  (1982). 
\bibitem{dmq} T. Damour, G.W. Gibbons, C. Gundlach, 
Phys. Rev. Lett. 64, 123 (1990); G. Piccinelli, F. Lucchin, 
S. Matarrese, Phys. Lett. B 277, 58 (1992).
\bibitem{HW1} J.C. Hwang, ApJ  375, 443 (1991);
 J.C. Hwang, Phys. Rev. D 53, 762 (1996). 
\bibitem{HW2} J.C. Hwang, Class. Quantum Grav. 7, 1613  (1990).   
\bibitem{MB} C.P. Ma, E. Bertschinger, ApJ 455, 7 (1995).
\bibitem{Gbounds} G.T. Gillies, Rep. Prog. Phys. 60, 151 (1997).
\bibitem{omega500} C.M. Will, Phys. Rep. 113, 345 (1984); 
T. Damour, to appear in Nucl. Phys. B (Proc. Suppl.) 1999; 
T. Damour, Eur. Phys. J. C3, 113 (1998);   
T. Damour,  in Les Arcs 1990, Proceedings,"New and exotic phenomena '90",
285 (1990); R.D. Reasenberg {\it et al.}, ApJ 234, L219 (1979).
\bibitem{faraoni} V. Faraoni, E. Gunzig, P. Nardone, 
Fund. Cosm. Phys., in press, preprint gr-qc/9811047
\bibitem{SM} J. Ellis {\it et al.}, Phys. Lett. B 134, 429 (1984);
E. Witten, Phys. Lett. B 155, 151 (1985); 
H. Nishino, E. Sezgin, Phys. Lett. B 144, 187 (1984).
\bibitem{bine} P. Bin\'etruy, preprint hep-ph/9810553.
\bibitem{sqcd} T.R. Taylor, G. Veneziano, S. Yankielowicz, 
Nucl. Phys. B 218, 493 (1983); I. Affleck, M. Dine, N. Seiberg, 
Phys. Rev. Lett. 51, 1026 (1983), Nucl. Phys. B 241, 493 (1984); 
A. Masiero, M. Pietroni, F. Rosati, preprint hep-ph/9905346.
\bibitem{SZ} U. Seljak,  M.  Zaldarriaga, 
ApJ 469, 437  (1996). 
\bibitem{liddle} A.R. Liddle, A. Mazumdar \& J.D. Barrow, 
preprint SUSSEX-AST 98/2-2, astro-ph/9802133 (1998)
\bibitem{SARKAR} S. Sarkar, Rep. Prog. Phys. 59, 1493 (1996).
\bibitem{eqofstate} G. Efstathiou, preprint astro-ph/9904356.
\bibitem{HSWZ} W. Hu, U. Seljak, M. White, M. Zaldarriaga,
Phys. Rev. D 57, 3290 (1998).
\bibitem{sigma8} L. Wang, P.J. Steinhardt, ApJ 508, 483 (1998).
\bibitem{CMBFUTURE} MAP home page: http://map.gsfc.nasa.gov/ ;
Planck Surveyor home page:
http://astro.estec.esa.nl/SA-general/Projects/Planck/
\bibitem{LSSFUTURE} SDSS home page:
 http://www.sdss.org/  ;
2dF home page: http://msowww.anu.edu.au/~colless/2dF/
\bibitem{kamion} X. Chen, M. Kamionkowski, preprint 
astro-ph/9905368. 
\end{thebibliography}
\end{document}